\documentclass[reprint, amsmath, amssymb, aps]{revtex4-2}

\usepackage{graphicx}
\usepackage{dcolumn}
\usepackage{bm}
\usepackage{xcolor}
\usepackage{version}
\usepackage{tikz}
\usetikzlibrary{shapes.geometric}
\usepackage{amsmath}
\usepackage{appendix}
\usepackage{hyperref}
\hypersetup{
    colorlinks = true,
    allcolors = {blue},
}

\begin{document}

\preprint{APS/123-QED}

\title{Dynamical Triplet Unravelling: A quantum Monte Carlo algorithm for reversible dynamics}

\author{Romain Chessex}
\email{rchessex@ethz.ch}
\author{Massimo Borrelli}
\author{Hans Christian \"{O}ttinger}

\affiliation{Polymer Physics, Department of Materials, ETH Z\"{u}rich, CH-8093 Z\"{u}rich, Switzerland}
\date{\today}

\begin{abstract}
We introduce a quantum Monte Carlo method to simulate the reversible dynamics of correlated many-body systems. Our method is based on the Laplace transform of the time-evolution operator which, as opposed to most quantum Monte Carlo methods, makes it possible to access the dynamics at longer times. The Monte Carlo trajectories are realised through a piece-wise stochastic-deterministic reversible evolution where free dynamics is interspersed with two-process quantum jumps. The dynamical sign problem is bypassed via the so-called \textit{deadweight approximation}, which stabilizes the many-body phases at longer times. We benchmark our method by simulating spin excitation propagation in the XXZ model and dynamical confinement in the quantum Ising chain, and show how to extract dynamical information from the Laplace representation.
\end{abstract}

\maketitle
\section{Introduction}
Recent years have witnessed an ever-growing interest in the many-body dynamics of strongly correlated quantum systems \cite{nonequil_nature, experiment_confinement}. Technical advancements in quantum state preparation, control and measurement have made it possible to observe the real-time evolution of typical systems in condensed matter physics at very low temperatures. Several experimental platforms, ranging from cold atoms in optical lattices \cite{experimental_heisenberg, bose_hubbard1,blochre2012,Grossre2017,schaferre2020}, to trapped ions \cite{blattrev2012,RevModPhys.93.025001} and superconducting circuits \cite{Romero2017,PhysRevB.98.174505}, have emerged as quantum simulators and different many-body Hamiltonians have been realized. The plethora of phenomena that have been investigated is very rich and includes many-body equilibration versus localization \cite{Alessio,ALET2018498,Deutsch_2018}, many-body dissipation, dynamics of quantum correlations \cite{cheneau2012,Geiger_2014,PhysRevLett.128.150602}, quenched many-body dynamics, and dynamical phase transitions \cite{dqpt,dqpt_review,PhysRevLett.125.143602}. 

Parallel to the abundance of experimental investigations, numerical methods \cite{real_time_review, eth_review, dqpt_review} have been developed to simulate condensed matter systems efficiently, including molecular and atomic gases in optical lattices. At the intersection between many-body physics and quantum information theory, tensor networks have been extensively explored in 1d systems, and nowadays, they are considered as a universal representation of low-energy, weakly entangled quantum states. Within this framework, time-dependent density matrix renormalization group \cite{tdmrg_review, tdmrg_transport1, tdmrg_transport2, tdmrg_impurity} has proven very effective in handling large particle systems and has established itself as a reliable standard tool. For higher dimensional systems and longer simulation times, nonequilibrium dynamical mean field theory \cite{review_dmft, mean_field_review} has been successfully applied to study spectroscopy \cite{ bose_hub_real_time_method}, nonlinear optics \cite{PhysRevB.62.R4769} and transport in solid-state physics \cite{PhysRevB.86.085110} as well as quenches \cite{eth2d} and relaxation dynamics in cold atoms in optical lattices \cite{PhysRevLett.100.120404,Eckstein_2010}. Finally, other computationally efficient methods have found a multitude of applications in quantum chemistry, with prominent examples being time-dependent Hartree-Fock \cite{tdhf}, multiconfigurational time-dependent Hartree-Fock \cite{mctdhf}, in which some electron-electron correlations are also included, and density functional theory \cite{RevModPhys.87.897}.

Most of the above techniques rely explicitly on symmetries in the Hamiltonian to cope with the large size of the Hilbert space. As opposed to deterministic techniques, quantum Monte Carlo methods (QMC) sample stochastic trajectories in a configuration space, onto which the original Hilbert space is directly mapped, and have proven very reliable in a number of different applications \cite{real_time_qmc, fciqmc_real_time, diagrammatic_qmc, hubbard_2d, hubbard2d_technique, RT_method_qft}. Although QMC methods were originally formulated to study ground state and equilibrium properties, dynamical adaptations have also been put forward that are based on stochastic sampling of diagrammatic expansions of many-body Green's functions defined along a Keldysh contour \cite{diagrammatic_qmc, diagrammatic_qmc2, diagrammatic}. All these methods, however, have been limited in accessing long-time behaviour by the well-known dynamical sign problem. Notable exceptions are the inchworm algorithm presented in \cite{impurity_sgn_prob} and real-time full configuration interaction QMC algorithm \cite{fciqmc_real_time}. The first technique can access medium-to-long-time $t\sim10$ dynamical properties, due to corrections of short-time diagrams, whereas the second can allow for simulation times up to $t\sim 40$, although at the expense of probability conservation.

Here, we present a new QMC algorithm that allows us to simulate the reversible dynamics of a quantum many-body system at long times. In some technical aspects, this method draws from the fixed point quantum Monte Carlo method (FPQMC), which was first introduced in \cite{fpqmc} for ground state calculations. The reversible dynamics, as dictated by the von Neumann equation, is reformulated in the Laplace representation and unraveled by stochastic trajectories. These are generated by a piece-wise deterministic-stochastic evolution where the free part of the many-body Hamiltonian is solved exactly, whilst particle-particle interactions are simulated using two-process quantum jumps. The QMC walkers sampling the configuration space are called triplets and consist of two state vectors, and a statistical weight. Depending on this weight, a triplet might spawn a new triplet via a two-process quantum jump or be removed from the simulation altogether. Two key ingredients are introduced. First, the \textit{deadweight approximation}, which stabilizes the phases originating from fast oscillating exponentials that are responsible for the dynamical sign problem. Second, as previously formulated in FPQMC and relying on \textit{dynamic norm}, importance sampling guarantees fast convergence of the time-dependent quantities under scrutiny. Near-exact trace conservation of the density matrix is guaranteed for the whole duration of the simulation.
We benchmark our algorithm against two case-studies on quantum lattices i) dynamics of a spin excitation following a quench in the Heisenberg XXZ model and ii) excitation confinement in the quantum Ising model \cite{confinement_nature, string_breaking,confinement}. In both cases we find results in excellent agreement with previous established literature. 

This manuscript is organized as follows; in Sec.\ \ref{tm_op}, we present the theoretical background for our numerical method and introduce some physical quantities of interest. In Sec.\ \ref{algo}, we break down the algorithm itself and explain in detail all its major components. In Sec.\ \ref{results}, we test our method for the case studies mentioned earlier and analyze the results. Finally, we draw some conclusion and illustrate some open perspectives in Sec.\ \ref{concs}.

\section{Theoretical basis}\label{tm_op}
\subsection{Dynamics in the Laplace space}
We focus on many-body systems whose Hamiltonian $H$ can be split into a free part and an interacting part, that is $H = H^{\text{free}} + H^{\text{int}}$. The adjectives free and interacting do not necessarily refer to the actual kinetic energy versus interactions within the Hamiltonian, but rather to whether the exact eigenstates are known \textit{a priori}, which is assumed to be always the case for $H^{\text{free}}$. The time evolution is dictated by the \textit{von Neumann} equation
\begin{equation}
    \frac{d}{dt}\rho = \mathcal{L}\rho\equiv -i\left[H,\rho\right],
    \label{eq:vonneuman}
\end{equation}
where $[\cdot,\cdot]$ denotes the commutator and $\rho$ the density matrix of the many-body system. If the total Hamiltonian is time independent, the formal solution of Eq. (\ref{eq:vonneuman}) reads
\begin{equation}
    \rho_t = \mathcal{E}_t \rho_{0}= e^{t\mathcal{L}}\rho_0,
\end{equation}
for some initial state $\rho_0$. To use the framework developed in \cite{fpqmc}, we transform the time-evolution superoperator $\mathcal{E}_t$ in the Laplace domain, that is
\begin{equation}
    \mathcal{R}_s = \int_0^\infty \mathcal{E}_t e^{-st} \text{d}t.
    \label{eq:laplace}
\end{equation} 
We define the superoperators
\begin{equation}
    \mathcal{L}^{\text{free}}\rho = -i[H^{\text{free}}, \rho], \qquad        \mathcal{L}^{\text{int}}\rho = -i[H^{\text{int}}, \rho],
\end{equation}
generating the dynamics described by the superoperators $\mathcal{R}^{\text{free}}_s$ and $\mathcal{R}^{\text{int}}_s$, respectively. By first Laplace transforming the formal solution of Eq.~\eqref{eq:vonneuman} and then applying the geometric series (see \cite{hcoQFT} for details), one arrives at the following expression for the Laplace transform of the total evolution superoperator $\mathcal{R}_{s}$
\begin{equation}
    \mathcal{R}_s =  \sum_{m = 0}^\infty r^m\left[\mathcal{R}_{s+r}^{\text{free}}\left(1 + \frac{\mathcal{L}^{\text{int}}}{r}\right)\right]^m\mathcal{R}_{s+r}^{\text{free}},
    \label{eq:magfor}
\end{equation}
with $r > 0$. We refer to Eq.\ \eqref{eq:magfor} as the \textit{magical formula} because it unifies a perturbative expansion and a numerical scheme in a single equation. If $r = 0$, the perturbative expansion is recovered, since the $m^{\text{th}}$ term in the sum corresponds to the $m^{\text{th}}$ order in perturbation theory. Conversely, if $r > 0$, the sum can be interpreted as the time-evolution operator of a numerical integration scheme with a $1/r$ time-step. Hence, truncating the magical formula at some $M_{\text{trunc}}$ term sets a natural time limit for the numerical integration, which is $t_{\text{max}}\sim \frac{1}{s_{\text{min}}} \sim \frac{M_{\text{trunc}}}{r}$. 

The Laplace transform of the time-evolution superoperator \eqref{eq:laplace} does not guarantee trace conservation by itself. However, noting that
\begin{equation}
    \text{Tr}(\mathcal{R}_s\rho_0) = \int_0^{\infty}\text{Tr}(\rho_t)e^{-st}\text{d} t = \frac{1}{s},
\end{equation}
one recovers the correct normalization for the quantity $\tilde{\rho}_s$ defined by
\begin{equation}
    \tilde{\rho}_s = s \mathcal{R}_s\rho_0.
\end{equation}
Even though $\tilde{\rho}_s$ possess all properties of a density matrix, we have to be careful when interpreting it as a physical density matrix because it is constructed from an integration over the whole time domain. In the limit $s\to\infty$, the density matrix is associated to a physical state, however for any finite value $s$, $\tilde{\rho}_s$ can be interpreted only as a formal density matrix.  

Once the density matrix $\tilde{\rho}_s$ is obtained, relevant quantities of interest can be calculated. For instance, the $s$-dependent correlation functions read
\begin{equation}
    C_s^{AB} = \text{Tr}\left(AB\tilde{\rho}_{s}\right) = s\text{Tr}\left(AB\mathcal{R}_{s}\rho_{0}\right),
    \label{eq:corrfunc}
\end{equation}
where $A$ and $B$ are generic self-adjoint operators. 

\subsection{Two-process stochastic unravelling}
As mentioned earlier, the magical formula \eqref{eq:magfor} can be interpreted as an integration scheme for any $r>0$. In this respect, the superoperator
\begin{equation}
T_{r}(s) = r\left[\mathcal{R}_{s+r}^{\text{free}}\left(1 + \frac{\mathcal{L}^{\text{int}}}{r}\right)\right],
\label{eq:t_op}
\end{equation}
should be regarded as the fundamental propagator in the Laplace domain. As anticipated earlier, a truncation order $M_{\text{trunc}}$ is set in the numerical implementation, which leads to the following approximation of the evolution superoperator
\begin{equation}
    \mathcal{R}_s \approx  \sum_{m = 0}^{M_\textrm{trunc}} \left[T_{r}(s)\right]^{m}\mathcal{R}_{s+r}^{\text{free}}.
    \label{eq:trunc}
\end{equation}
From this approximation, we develop a stochastic process to unravel the von Neumann equation. We take inspiration from the method originally introduced as \textit{triplet unravelling} in \cite{fpqmc}. Our method tracks the evolution of triplets of the form $(c_{m},|\phi_{m}\rangle,|\psi_{m}\rangle)$ in the many-body Hilbert space where the subscript ${m}$ refers to the $m$-th order iterative term in the evolution described by the propagator \eqref{eq:t_op}. The evolution of the piece-wise-deterministic stochastic processes $|\phi_{m}\rangle$, $|\psi_{m}\rangle$ alternates between an exact continuous evolution, governed by $H^{\text{free}}$, and stochastic state jumps accounting for $H^{\text{int}}$. We construct these processes as to reproduce the application of the magical formula Eq.\ \eqref{eq:magfor} for infinite order of iteration $M_{\text{trunc}}\to \infty$ via
\begin{equation}
    \tilde{\rho}_s =  s \sum_{m = 0}^{\infty}\mathbb{E}[c_m|\phi_m\rangle\langle \psi_m|],
    \label{eq:rhos_unravelled}
\end{equation}
where $\mathbb{E}[\cdot]$ represents the expectation value of the stochastic processes. This construction ensures that the solution of the von Neumann equation in the Laplace domain is recovered in the limit of a large order of iteration. Because eigenstates of $H^{\text{free}}$ are assumed to be known exactly, we use these states as the computational basis allowing us to compute the free evolution exactly. In order to minimize the statistical correlations between the processes $|\phi_m\rangle$ and $|\psi_m\rangle$, we use two-sided jump processes and model the effect of $H^{\text{int}}$ as follows
\begin{equation}
    |\phi_m\rangle\langle \psi_m| \mapsto |\phi_m\rangle\langle \psi_m| - \frac{i}{r}(H^{\text{int}}|\phi_m\rangle\langle \psi_m| - |\phi_m\rangle\langle \psi_m|H^{\text{int}}),
\end{equation}
which coincides with a stochastic application of the superoperator $\left(1 + \frac{\mathcal{L}^{\text{int}}}{r}\right)$.

\section{Dynamic algorithm}
\label{algo}
The present algorithm is the dynamical adaptation of the FPQMC method introduced in \cite{fpqmc}, of which it preserves the overall structure. The latter consists of a main loop, whose application is iterated over an ensemble of walkers until a solution emerges. Two distinct steps compose the main loop, the \textit{interaction} and the \textit{free evolution} that are implemented by the superoperators $\left(1 + \frac{\mathcal{L}^{\text{int}}}{r}\right)$ and $\mathcal{R}_{s+r}^{\text{free}}$, respectively.

\subsection{The main loop}
The algorithm considers the evolution of an ensemble of Monte Carlo walkers, called triplets, defined as $\{(w_n,|i_n\rangle,|j_n\rangle)\}_n$, where $w_n$ is the (complex) weight factor and $|i_n\rangle,|j_n\rangle$ are the free Hamiltonian eigenstates. Here the subscript $n$ refers to the index in the triplet ensemble. In what follows, we will use the lighter notation $|i\rangle\equiv i$. As anticipated earlier, during the evolution, the triplets repeatedly experience stochastic spawning events, realized by state jumps associated with $H^{\textrm{int}}$, interspersed with continuous weight updates coming from the exact, free evolution $H^{\textrm{free}}$. The rate of the spawning events is $r$ and it is fixed prior to the simulation. 

In order to control the evolution of the ensemble size, we introduce the following parameters: i)  the walker's unit of weight $w_u$, which also fixes the initial walker's population of the statistical ensemble, ii) the dead weight $u_{\text{dw}}$ which is used throughout the simulation to identify statistically unimportant triplets, and iii) the spring constant $\kappa$ of the importance sampling procedure. The dead weight and spring constant are approximation and bias parameters, respectively. As such, they are not rooted in the theoretical framework introduced in Sec.~\ref{tm_op}, but they are extra computational features that are necessary for numerical stabilization and convergence. Both deadweight approximation and importance sampling are discussed in details in subsection \ref{subsec:dw_&_is}. However, as they are crucial for the execution of the main loop described in this subsection, we briefly recall why they are needed. The deadweight approximation aims to reduce the effect of the dynamical sign problem which is known to cause numerical instabilities connected to fast oscillating exponentials, especially at longer times. Importance sampling reduces the number of triplets needed to obtain a convergent simulation by preventing unwanted extra exploration of the Hilbert space. During the simulation, the walker's population is controlled by both $\kappa$ and $u_{\text{dw}}$. In general, if the spring constant is too large, the triplet population decreases fast. The same being true for the deadweight parameter, both parameters should be optimized. In principle, one should first find the smallest $u_{\text{dw}}$ that overcomes the dynamical sign problem and, subsequently, decrease the spring constant until a target value of the triplet's population is achieved. 

We now discuss in detail the main loop of our algorithm, depicted in Fig.~\ref{fig:finalflowchart}. The compression and decompression steps are thoroughly described in Appendix ~\ref{sec:comp}.
\begin{enumerate}
    \item \textit{\underline{Spawnings}}  
    \begin{enumerate}
        \item Preparation to interaction. 
            \begin{enumerate}
                \item Deactivation of unimportant triplets $(w_n,i_n,j_n)$. If $|w_n| < u_{\text{dw}}$, the triplet becomes inactive and survives with probability $|w_n|/u_{\text{dw}}$. The weight of surviving triplets is updated to $u_{\text{dw}}w_n/|w_n|$. Surviving inactive triplets do not experience any collision and are therefore evolved only freely.
                \item Pre-spawning decompression. The active triplets are split into $N_c = \lfloor |w_n|/w_{u}\rfloor$ child triplets with weight $w_n/N_c$. 
            \end{enumerate}
        \item Spawning. For each child triplet, a side is randomly chosen (quantum mechanically, either $i_n$ or $j_n$) and a new state $k_n$ is selected with equal probability among the $n_t$ possible spawning transitions. For instance, if the ket is chosen, the following triplet is added to the ensemble
            \begin{equation}
                (i\frac{H^{\text{int}}_{i_nk_n}}{r }\frac{w_n}{N_c}2n_tT_b, k_n, j_n),
            \end{equation}
            where $T_b = e^{\kappa(n_{ij}^2-n_{kj}^2)/2}$ is the transition bias, with $n_{ij}$ the minimum number of application of $H^{\text{int}}$ to transition from state $i$ to $j$ (see next subsection for details). If $j_n$ is selected, an equivalent triplet is spawned on the bra side.
    \end{enumerate} 
    \item \textit{\underline{Free evolution}}
        \begin{enumerate}
            \item Full compression. A class containing all the triplets associated to a specific pair $(i_n,j_n)$ is replaced by a single triplet whose weight is the sum of all the weights in the class.
            \item Weight update. For each class $(w_n,i_n,j_n)$, a weight update is performed, accounting for the free evolution
            \begin{equation}
                w_n \mapsto \frac{r}{s+r + i(H_{i_ni_n}^{\text{free}} - H_{j_nj_n}^{\text{free}})}w_n.
            \end{equation}
        \end{enumerate}
\end{enumerate}

In a typical simulation, correlation functions of the form of Eq.\ \eqref{eq:corrfunc} are calculated via a procedure involving an initialization, an iteration on the main loop, and a computation of correlation functions. After state initialization, which matches the triplet ensemble to the desired initial state $\rho_{0}$, the main loop is executed $M_{\text{trunc}}$ times, with the cutoff $M_{\text{trunc}}$ being set prior to the simulation. At the end of each loop $m\leq M_{\text{trunc}}$, the quantity
\begin{equation}
    \text{Tr}\left(AB\tilde{\rho}_s^{(m)}\right)
    \label{eq:num_measure}
\end{equation}
is calculated and stored, with $\tilde{\rho}_s^{(m)}$ being the matrix representation of the output of the $m-1$ loop. Finally, the correlation function can be computed by summing over all the contribution from all loops, i.e.
\begin{equation}
    C_s^{AB} \approx \sum_{m = 1}^{M_{\text{trunc}}} \text{Tr}\left(AB\tilde{\rho}_s^{(m)}\right).
\end{equation}

Due to the importance sampling procedure,  the ensemble is biased by the spring constant $\kappa$, and triplets' weight have to be compensated to obtain unbiased physical results. For a general operator $X$, physical expectation value is computed using the physical weights $e^{\frac{\kappa}{2}n_{ij}^2}w_n$ as follows
\begin{equation}
    \text{Tr}\left(X\tilde{\rho}\right) \approx \sum_{n}w_ne^{\frac{\kappa}{2}n_{ij}^2} X_{i_nj_n}, 
\end{equation}
where $\tilde{\rho}$ is represented by the ensemble $\{w_n,i_n,j_n\}_n$ and $X_{i_nj_n} = \langle i_n|X|j_n\rangle$.

\begin{figure}[h!]
\centering
    \includegraphics[width = \linewidth]{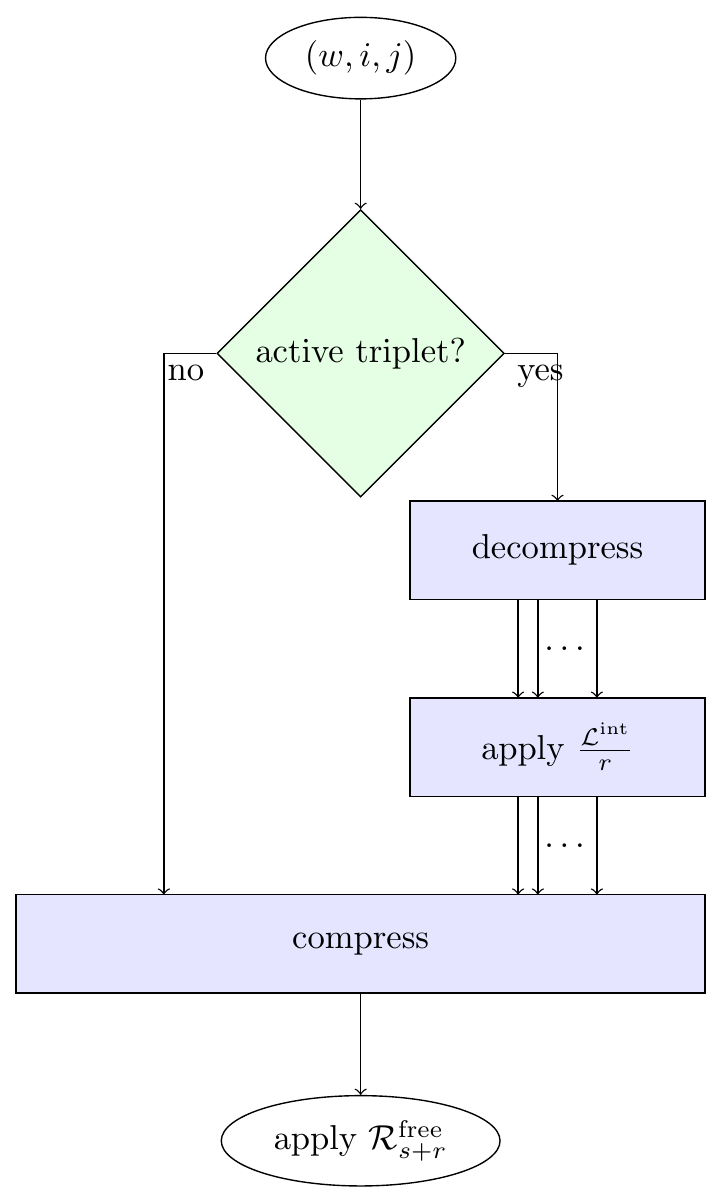}
    \caption{Flowchart of the algorithm describing the application of the superoperator $\mathcal{L}^{\text{int}}/r$. The green diamond represents the deadweight approximation.}
    \label{fig:finalflowchart}
\end{figure}

\subsection{Numerical stabilization and convergence}
\label{subsec:dw_&_is}
The size of a typical many-body Hilbert space imposes a strong demand on the computational resources and limits simulations to small systems. Furthermore, the dynamical sign problem represents a non-negligible technical issue in terms of numerical stability. In order to tackle these two problems, we complement the two-process unravelling loop described above with two extra procedures, importance sampling and the deadweight approximation. 

The importance sampling procedure relies on two main concepts, originally introduced and explained in details in Sec.\ III.C of \cite{fpqmc}. The first is the introduction of a \textit{dynamic norm} $n_{ij}$, a distance between states in the Hilbert space, defined as the minimum number of applications of $H^{\text{int}}$ to transition from state $i$ to $j$. The second concept associates to each triplet two weights, a physical weight $c$ and an ensemble weight $w$.  The physical weight $c$ is used to compute expectation values, {\itshape e.g.} $\text{Tr}(A\tilde{\rho}_s)$, whereas the ensemble weight $w$ reflects the number of jumps performed by the triplet. The latter weight allows to rate the statistical relevance of the triplet within its ensemble without altering the physical averages. These two are related via the equation $c = bw$, by a norm-dependent bias $b \equiv b(n_{ij})$. A standard choice for $b$ is $b = \exp\left(\frac{\kappa}{2}n_{ij}^2\right)$, which is a harmonic interaction with spring constant $\kappa$ between the states $i$ and $j$ that aims at reducing the number of triplets with a large dynamic norm. 

The dynamical sign problem manifests itself as numerical instabilities occurring at medium times. In order to avoid divergences, we need to prevent statistically irrelevant triplet from over-contributing to the dynamics. The basic idea of the deadweight approximation is to forbid all the triplets with an ensemble weight $w$ below a threshold $u_{\text{dw}}$ from experiencing jumps while still allowing them to undergo free evolution. The triplets undergoing the jumps are called active triplets whereas the ones below the threshold $u_{\text{dw}}$ as referred to as inactive. Intuitively, one can imagine these inactive triplets as an effective environment whose free dynamics is needed in order to guarantee an ergodic exploration of the Hilbert space by the remaining active triplets. 

An extra computational feature which greatly improves the efficiency of the algorithm concerns the parametric nature of $s$ in the magical formula \eqref{eq:magfor}.  Unlike time evolution, where typically each iteration at a given physical time $t' = t + \Delta t$ produces an updated density matrix $\rho_{t'}$, the evolution dictated by the magical formula yields results for a density matrix $\tilde{\rho}_{s}$ that represents an integral over the whole time range, as reflected in the use of the Laplace transform ~\ \eqref{eq:laplace}. At a first glance, one might be skeptical about the efficiency of our method as one would need to produce an independent simulation for each $s$ value to reconstruct $\tilde{\rho}_{s}$. However, by noting that the $s$ variable only appears in the exact free part of the total evolution, one can produce results for the whole $s$ range in a single simulation at a minimal additional computational cost. For example, two triplets $(c^{(1)}_m ,|\phi^{(1)}_m \rangle, |\psi^{(1)}_m \rangle)$ and $(c^{(2)}_m ,|\phi^{(2)}_m \rangle, |\psi^{(2)}_m \rangle)$ having the very same trajectory for two different values $s_1\neq s_2$ differ from one another only in the application of the free evolution operator $\mathcal{R}_{s+r}^{\textrm{free}}$. The idea is then to evolve a single copy of these triplets and reweight each application of the free evolution operator when computing observables simultaneously for $s_1$ and $s_2$. In practice, we select the range of $s$ values we are interested in before the simulation, and at each iterative order $m$, we reweight the triplets for the whole range of $s$ values during the free evolution part.

\section{Results}
\label{results}
In this section, we test our algorithm on two 1d models in quantum magnetism, namely the Heisenberg XXZ and the Ising model, and look at the dynamics of excitations following a quench.
We stress that, being a QMC method, it can be applied to 2d and 3d systems as well, with the only extra difficulty being the increasing size of the Hilbert space.
Although the use of the Laplace domain indicates that the long-term quantities can be naturally computed, we will show that certain oscillation frequencies can also be extracted from functions computed on the Laplace domain. 

\subsection{Heisenberg XXZ model: analysis of the method and quenched dynamics}
In order to assess the efficiency of our method, we first present a quantitative analysis of our numerical method. In particular we focus on the walker's population control via the parameters $w_u$ and $u_{\text{dw}}$, the dependence of the end result on the importance sampling parameter $\kappa$ and the trace preserving properties of the algorithm. 

Like in any Monte Carlo method, the walker population is the main quantity responsible for the size of the statistical error bars. Within this method, the population for a given loop refers to the number of spawning attempts during that loop, and due to the high number of spawnings, it naturally increases at an exponential rate. It is therefore vital to slow down any possible overgrowth and limit the exploration of the Hilbert space to statistically important triplets only. In practice, whenever the target value of the triplets' population is reached, the deadweight approximation is enabled by setting a non-vanishing, fine-tuned threshold $u_{\text{dw}}$. This drastically reduces exponential increase in population and further keeps it at a nearly constant value. 

To be more concrete we consider the well-known Heisenberg XXZ model with $L$ spins and open boundary conditions. Its Hamiltonian reads
\begin{equation}
    H = J_{xy}\sum_{i = 1}^{L-1} (\sigma_i^x \sigma_{i+1}^x + \sigma_i^y \sigma_{i+1}^y) + J_z\sum_{i = 1}^{L-1} \sigma_i^z \sigma_{i+1}^z,
    \label{eq:xxz}
\end{equation}
where $\sigma^x, \sigma^z$ are the standard Pauli matrices and the sum runs over the spins in the chain. Following the recipe introduced earlier, the Hamiltonian~\eqref{eq:xxz} is split into a free and an interacting part, $H = H^{\text{free}} + H^{\text{int}}$, as follows
\begin{equation}
    H^{\text{free}} = J_z\sum_{i} \sigma_i^z \sigma_{i+1}^z, \; H^{\text{int}} =J_{xy}\sum_{i} \sigma_i^+\sigma^-_{i+1} + \sigma_{i}^-\sigma_{i+1}^+,
\end{equation}
where $\sigma^\pm = \sigma^x \pm i\sigma^y$. This particular splitting of the Hamiltonian sets the eigenbasis of $H^{\text{free}}$ as the computational basis, that is
\begin{equation}
    |e_{s_1,\dots,s_{L}}\rangle = |s_1\rangle \otimes |s_2\rangle \otimes \dots \otimes |s_{L}\rangle,
\end{equation}
where $s_a = s^z_a= \pm 1$. The single-particle operators $\sigma^z$ and $\sigma^\pm$ act of the basis states according to the standard algebra of Pauli matrices 
\begin{equation}
    \sigma^z|\pm 1\rangle = \pm|\pm 1\rangle, \qquad \sigma^\pm  |\mp 1\rangle = |\pm 1 \rangle, \qquad \sigma^\pm |\pm 1\rangle = 0.
\end{equation}
The initial state is set to
\begin{equation}
    \rho_{\text{ini}} = |\text{ini}\rangle\langle\text{ini}|,
\label{eq:inixxz}
\end{equation}
with 
\begin{equation}
    |\text{ini}\rangle = \bigotimes_{i = 1}^{L/2} |+1\rangle_i \bigotimes_{i = L/2 +1}^{L}|-1\rangle_i,
\end{equation}
in which strongly localized spin excitations serve as an initial source of energy. This initial state will be useful later to understand the propagation of spin excitations across the chain.

\begin{figure}
    \centering
    \includegraphics[width = \linewidth]{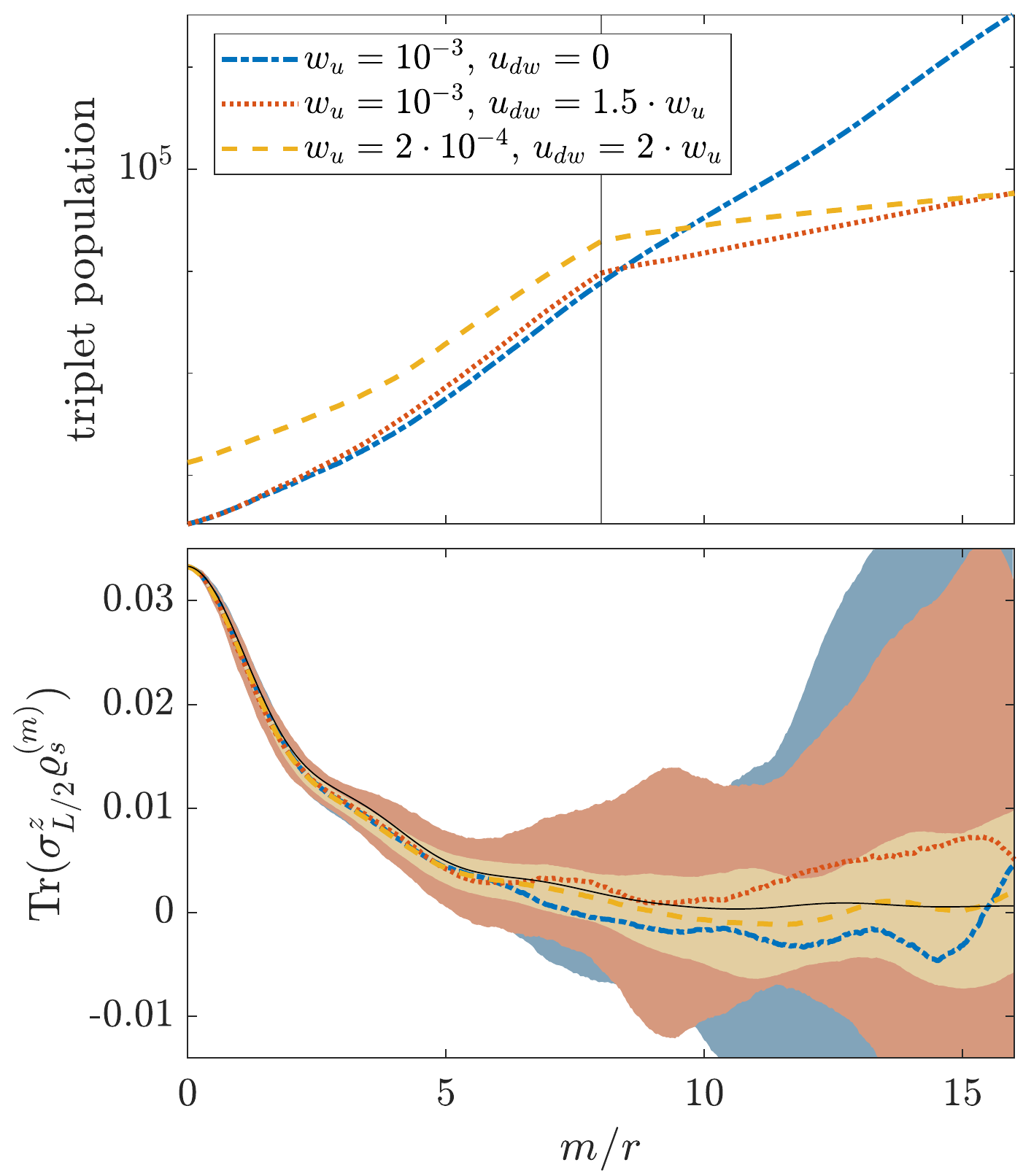}
    \caption{Upper panel: Evolution of the triplet population for various values of the deadweight threshold $u_{\text{dw}}$ and triplet's unit weight $w_u$ for the $L = 10$ XXZ model with $J_{xy}=1$, $J_z = 0.9$, $s=1/20$, $r = 30$ and $\kappa = 2$. The error bars were calculated over 30 independent simulations. Lower panel: evolution of the quantity $\text{Tr}(\sigma_{L/2}^z\tilde{\rho}^{(m)}_s)$ defined in Eq.\ (\ref{eq:num_measure}) for the same parameters as above.}
    \label{fig:popctrl}
\end{figure}

Figure \ref{fig:popctrl} shows three examples of the evolution of the population and the accuracy of the numerical integration for these populations in the $L=10$ XXZ model, for $J_z = 0.9$, $J_{xy} = 1$, $s=5\cdot 10^{-2}$, $r=30$ and $\kappa = 2$. In all the examples, the population control is performed by enabling the deadweight approximation at $m/r = 8$, represented by the vertical line in the upper panel. The value of the threshold $u_{\text{dw}}$ then controls the second phase of the exponential increase of the population. 
In passing, we note that $w_u$ can be responsible for some secondary effects later during the evolution, \textit{e.g.}, the decreasing distance between the dotted red and dashed orange lines in the first stage in Fig.\ \ref{fig:popctrl}.

\begin{figure}
    \centering
    \includegraphics[width = \linewidth]{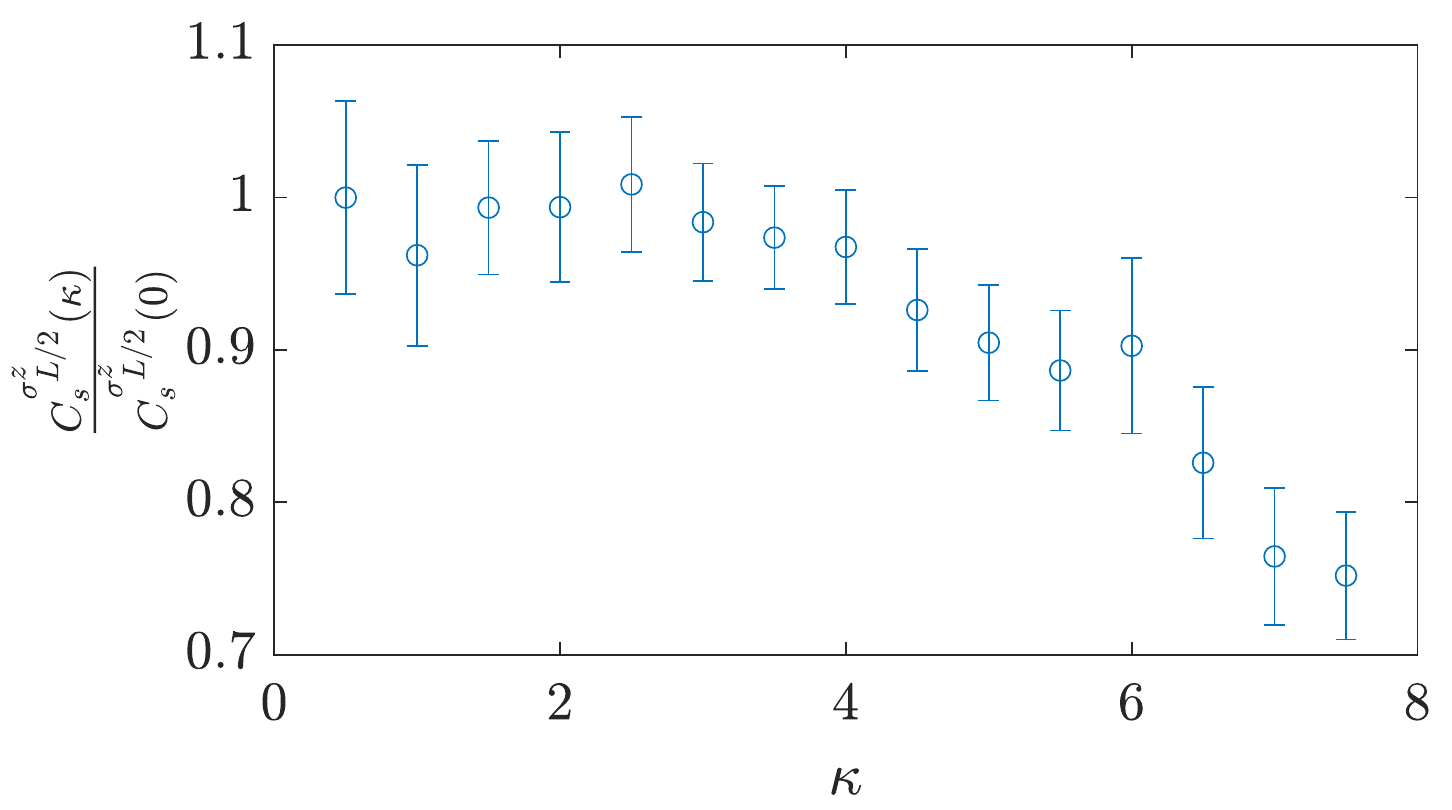}
    \caption{Dependence of the spring constant $\kappa$ on the final result $C_s^{\sigma^z_{L/2}}(\kappa)$ defined in Eq.\ \eqref{eq:corrfunc} for a $L =8$ XXZ model with $J_z = 1.5$, $J_{xy} = 1$ and $s=1/10$. The errors were calculated over 100 independent simulations with $u_{\text{dw}}=1.125 w_u$ and $r = 30$.}
    \label{fig:kappa_dep}
\end{figure}

The introduction of the importance sampling procedure was motivated by the need to reduce the number of triplets in the ensemble to only the statistically important ones. Thus, the main effect of the $\kappa$ parameter is a reduction of the number triplets with a large dynamic norm. A larger $\kappa$ leads to a smaller exponential increase of the population. However, if the spring constant is chosen too large, it may result in an underestimate of the final values of the correlation functions \eqref{eq:corrfunc}. This gap originates from breaking the ergodicity of Hilbert space exploration, with the triplets being confined to a small region of the Hilbert space around their initial state. Figure \ref{fig:kappa_dep} illustrates the underestimation of the quantity $C_s^{\sigma_{L/2}^z}$ in the $L=8$ XXZ model with $J_z =1.5$, $J_{xy} = 1$ for different values of the spring constant. We note that the importance sampling induces a systematic error for values $\kappa\leq4$, a critical value below which the population is large enough to avoid ergodicity breaking. On a similar note, we remark that the deadweight approximation can also affect the final results and a $u_{\text{dw}}$ analysis should be carried out, similarly to $\kappa$. In order to obtain the optimal threshold, it is more efficient to start from a larger value of $u_{\text{dw}}$ and further decrease it until the final result becomes constant within the acceptable statistical error. 

\begin{figure}
    \centering
    \includegraphics[width =  \linewidth]{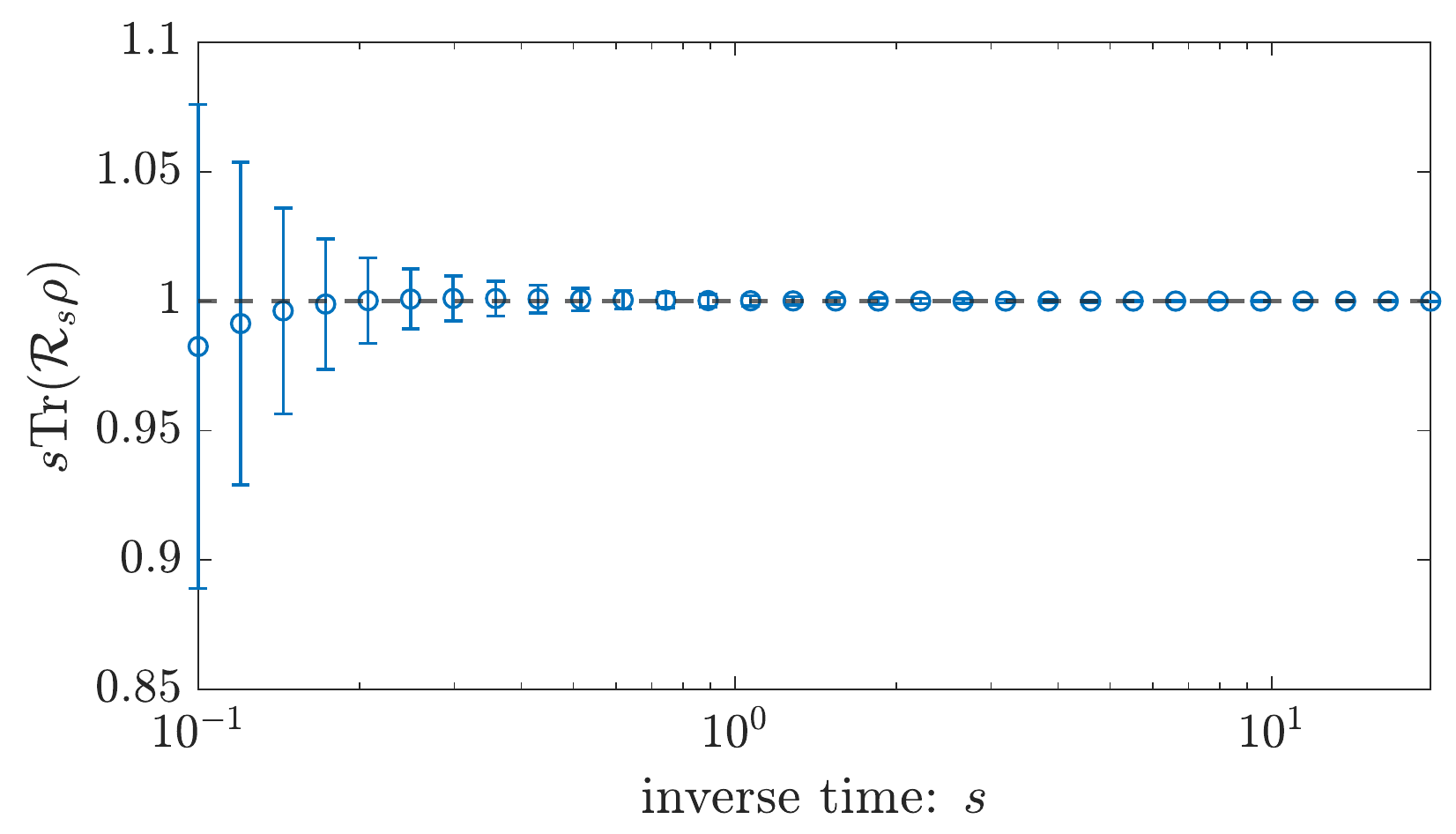}
    \caption{Trace conservation of the magical formula Eq.\ \eqref{eq:magfor} for the $L=20$ XXZ model with $J_z = J_{xy} = 1$. The population was about $2\cdot 10^4$ triplets with $r=100$, $\kappa = 4$, $u_{\text{dw}} = 1$ at $m/r=4$ and the error bars were calculated over 30 independent simulations.}
    \label{fig:unitarity}
\end{figure}

The trace conserving character of the evolution operator in the Laplace domain $\mathcal{R}_{s}$ can be checked using the following condition \cite{hcoQFT}
\begin{equation}
    \lim_{s\to 0} \text{Tr}\left(\tilde{\rho}_s\right) = \text{Tr}(\rho_0).
    \label{eq:unitary}
\end{equation}
If the algorithm violates the unitary character of the time-evolution, instabilities will occur and grow exponentially with the simulation length. Figure \ref{fig:unitarity} shows trace conservation according to Eq.\ \eqref{eq:unitary} with $\rho_0$ chosen as the ground state of the XXZ Hamiltonian with $L=20$. The exponential growth of the error bars for small $s$ signals the occurring of the dynamical sign problem. The severity the dynamical sign problem generally depends on the details of the system as well as the observables being computed.

\begin{figure}
    \centering
    \includegraphics[width = \linewidth]{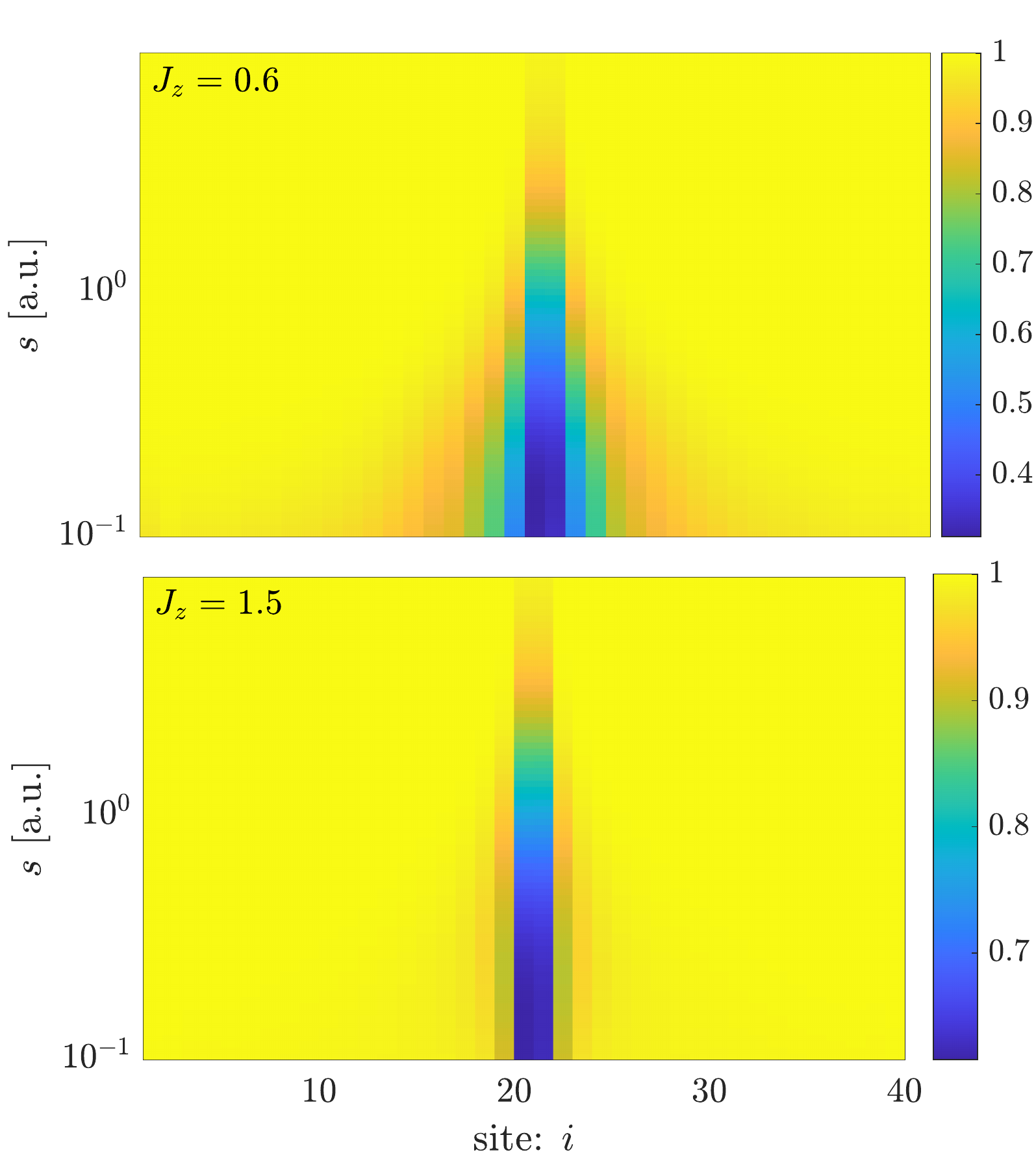}
    \caption{Evolution in the Laplace domain of the magnetization profile $|C_s^{\sigma^z_i}|$ for the $L = 40$ XXZ chain for $J_{xy} = 1$. Ballistic behaviour is clearly visible in the upper panel, whereas transport is almost completely suppressed in the lower panel. The values plotted were averaged over 100 independent simulations with a population of about $10^5$ triplets and $r = 100$, $\kappa = 4$, $u_{\text{dw}} = 1$ at $m/r = 12$.}
    \label{fig:magngrid}
\end{figure}

\begin{figure}
    \centering
    \includegraphics[width  = \linewidth]{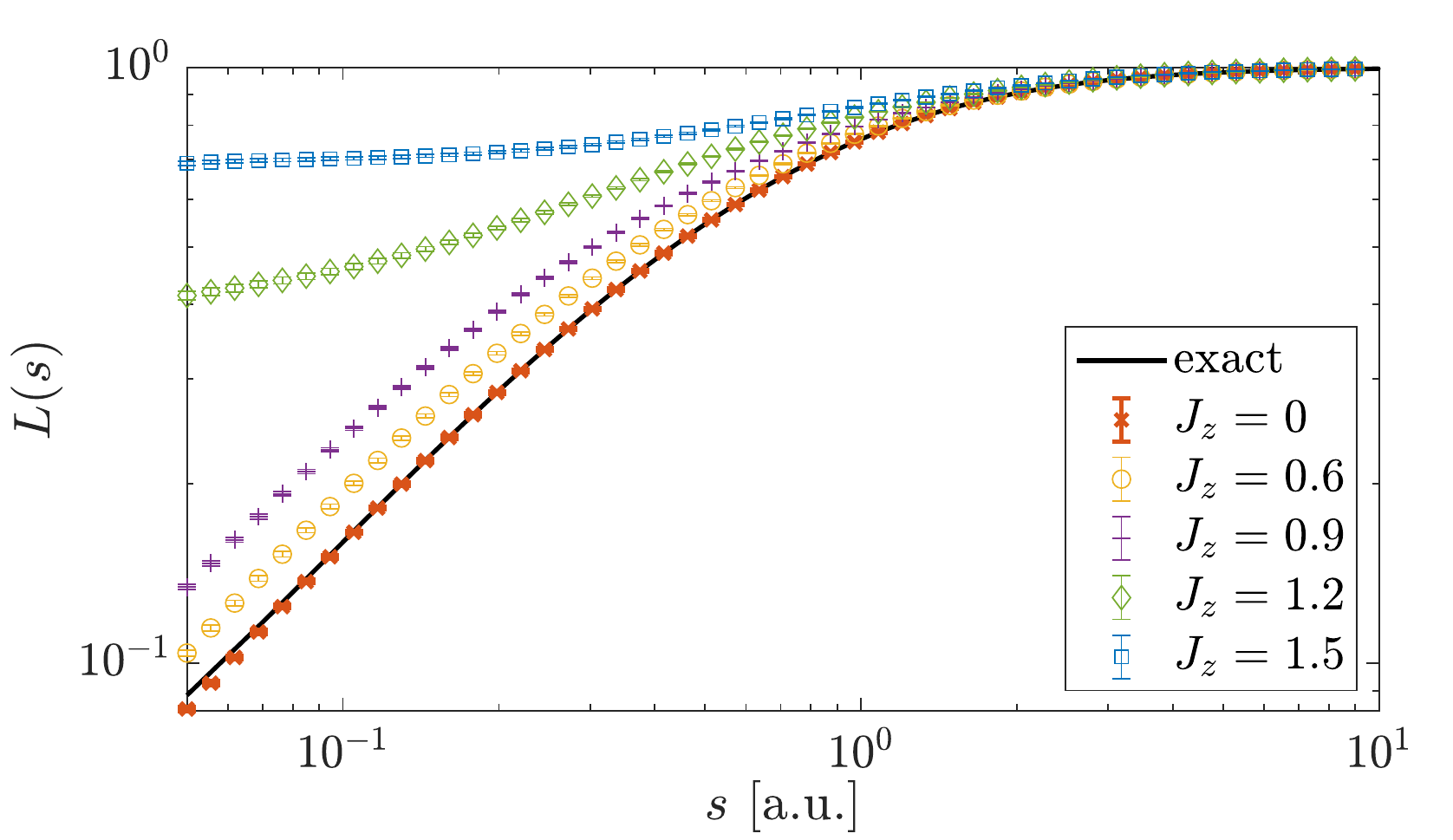}
    \caption{Evolution of the Loschmidt echo in the Laplace domain $L(s) = s\;\text{Tr}(\rho_{\text{ini}}\mathcal{R}_s\rho_{\text{ini}})$
 for the $L=40$ XXZ chain with $J_{yx}=1$ and various values of $J_z$. The errors were estimated over 30 independent simulations, with $r = 30$, $\kappa = 1$ and $u_{\text{dw}} = 1.2$ at $m/r=12$. The population at the end of each simulation was about $2\cdot 10^6$ triplets. The continuous curve represents the exact solution for $J_z = 0$.}
    \label{fig:loschmidt}
\end{figure}

We now focus on the dynamical behaviour of spin excitations following an instantaneous quench. For $J_z <J_{xy}$ these travel ballistically across the chain regardless of its length, whilst, on the contrary, for $J_z >J_{xy}$ propagation is inhibited by the strong $\sigma_{z}$ coupling \cite{tdmrg_transport1}. The propagation of these spin excitations can be quantified by analyzing the evolution of the magnetization profile. Starting from $\rho_{\text{ini}}$, this profile is calculated as
\begin{equation}
    C_s^{\sigma_i^z} \equiv \;\text{Tr}( \sigma_i^z\tilde{\rho}_s).
    \label{eq:osm}
\end{equation} 
Figure \ref{fig:magngrid} illustrates the evolution of the absolute value of the magnetization profile \eqref{eq:osm} for $J_{xy}=1$ in both the ballistic and strongly interacting regimes at $J_z = 0.6$ (top) and $J_z = 1.5$ (bottom), respectively. In the upper panel, a power-law decay can be observed. This is consistent with ballistic propagation where the spin excitations travel within a light-cone type of spatial and temporal region. On the contrary, in the lower panel at $J_z = 1.5$ magnetization propagation is practically suppressed as a result of the localizing longitudinal interactions.
To measure this sharp transition between ballistic and suppressed propagation, we consider the Loschmidt echo $L(t) = |\langle \psi_0|e^{-iHt}|\psi_0\rangle|^2$, a measure of the disturbance induced on a quantum system by an external perturbation. In the Laplace domain, this reads, 
\begin{equation}
    L(s) = s\;\text{Tr}( \rho_0 \mathcal{R}_s\rho_0) \equiv \text{Tr}( \rho_0 \tilde{\rho}_s),
\end{equation}
where $\rho_0 = |\psi_0\rangle \langle \psi_0|$ is the initial state. Figure \ref{fig:loschmidt} illustrates the evolution of $L(s)$ for $J_{xy}=1$ at various $J_z$ for a spin chain of length $L=40$ from the initial state Eq.\ \eqref{eq:inixxz}. Increasing $J_z$ in the Hamiltonian \eqref{eq:xxz} suppresses the propagation of magnetization. The quantum state of the spin chain essentially freezes up as a consequence of strong localization, and this results in a weak decay to nearly constant Loschmidt echo. This observation is supported by previous results obtained via exact diagonalization \cite{XXZ_quenches}. 

\subsection{Confinement in the Quantum Ising model}
For our second case study, we consider a ferromagnetic quantum Ising chain of length $L$ coupled to the transverse and longitudinal magnetic fields $h_x$ and $h_z$, respectively. The Hamiltonian reads 
\begin{equation}
    H = -J \sum_{i = 1}^{L-1} \sigma_i^z\sigma_{i+1}^z - h_z \sum_{i = 1}^L\sigma_i^z - h_x \sum_{i =1}^L\sigma_i^x,
\end{equation}
where $\sigma^x, \sigma^z$ are the standard Pauli matrices and the sum runs over the spins in the chain. In the absence of a longitudinal field, the model is exactly solvable and has a phase transition at $h_x = J$. In the ordered phase $h_x < J$, the ground state is degenerate due to a spin-flip symmetry breaking and corresponds to domain walls of various lengths. For instance, for $h_x = h_z =  0$, the domain walls' lengths are maximized, and the degenerate ground states are $|\Psi_{\textup{u}}\rangle = \bigotimes_{i = 1}^L|\uparrow\rangle_i$ and $|\Psi_{\textup{d}}\rangle = \bigotimes_{i = 1}^L|\downarrow\rangle_i$. A non-zero value of the longitudinal field $h_z$ creates an energy gap between these states by increasing the energy of the spin domains along the field $h_z$. Hence, the latter field acts as an attracting potential between the two walls delimiting a domain. 

\begin{figure}
    \centering
    \includegraphics[width = \linewidth]{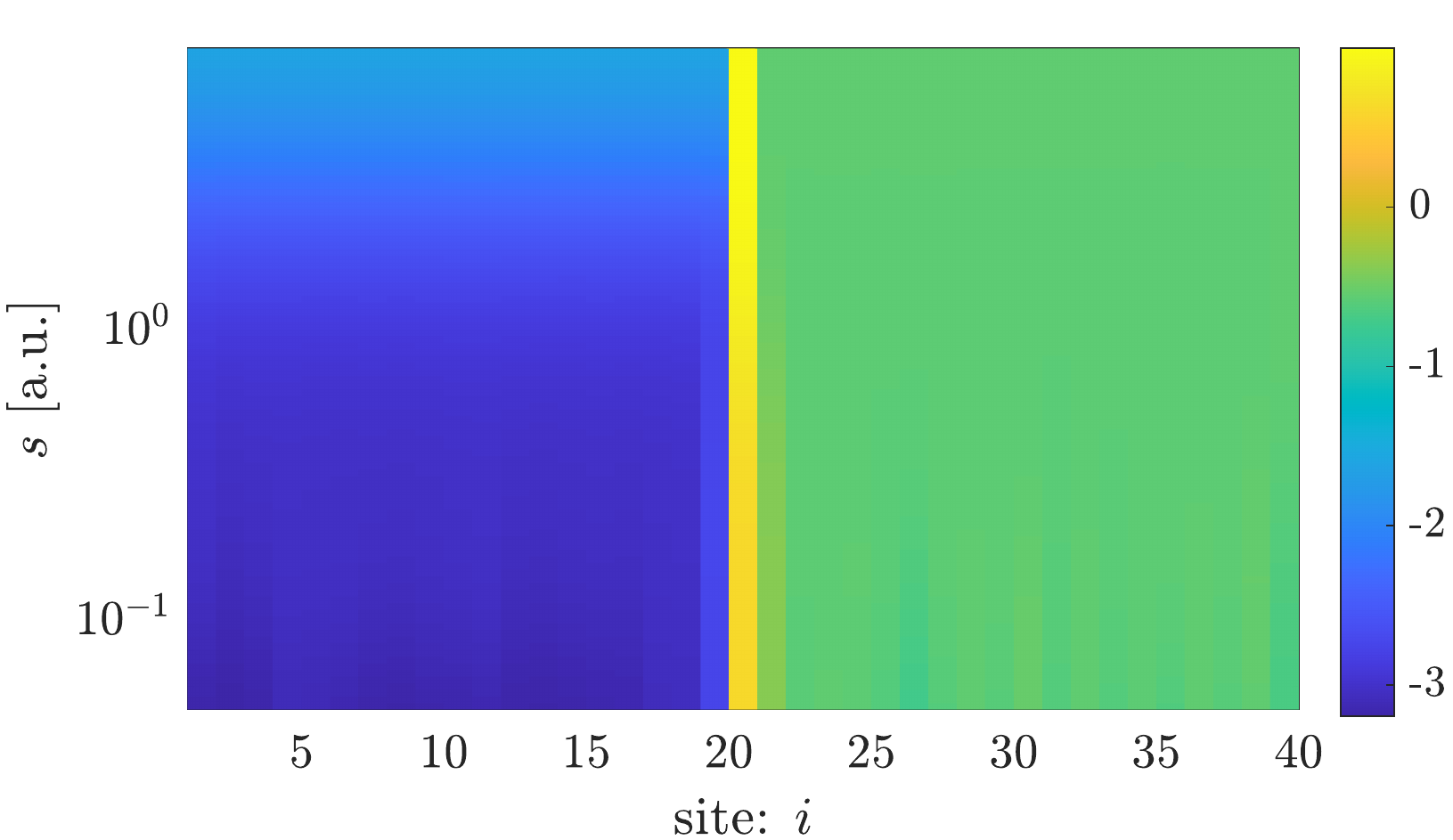}
    \caption{Evolution in the Laplace domain of the energy density profile $C_s^{H_i}$ for a $L=40$ quantum Ising chain with $J=1$, $h_x = 0.2$, and $h_z = 1.2$ starting from a state with a single kink in the middle. The profile was computed over 30 independent simulations for $r=30$, $\kappa = 3$ and $u_{\text{dw}} = 1$ at $m/r = 6$.}
    \label{fig:energydens}
\end{figure}

\begin{figure}
    \centering
    \includegraphics[width = \linewidth]{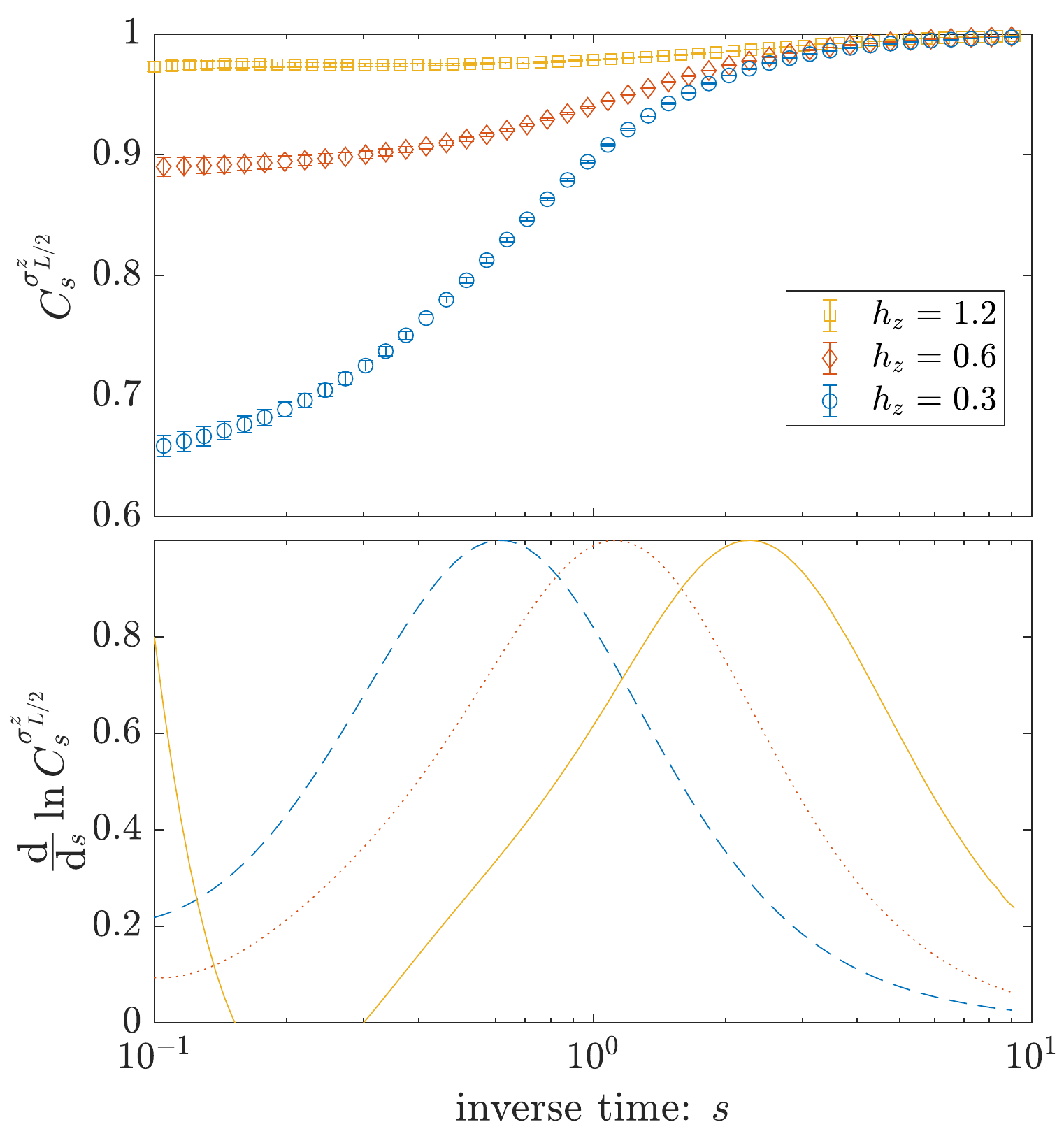}
    \caption{Upper panel: evolution in the Laplace domain of the mean magnetization of the spin on the left of the central opposite pair $C_s^{\sigma_{L/2}^z}$ for the $L=20$ quantum Ising chain with $J =1$, $h_x=0.2$ and the parameters $r=30$, $\kappa = 2$, and $w_u = 4\cdot 10^{-6}$. The dead weight approximation was enable at $m/r = 6$ with $u_{\text{dw}}=1.6,~1.5,~1.8$ for $h_z = 1.2,~0.6,~0.3$, respectively. The choice of specific thresholds $u_{\text{dw}}$ originates from the optmization procedure. Lower panel: logarithmic derivative of the magnetization $C_s^{\sigma_{L/2}^z}$. The derivative was calculated using a cubic spline fitting.}
    \label{fig:confinement_oscillations}
\end{figure}

\begin{figure}
    \centering
    \includegraphics[width = \linewidth]{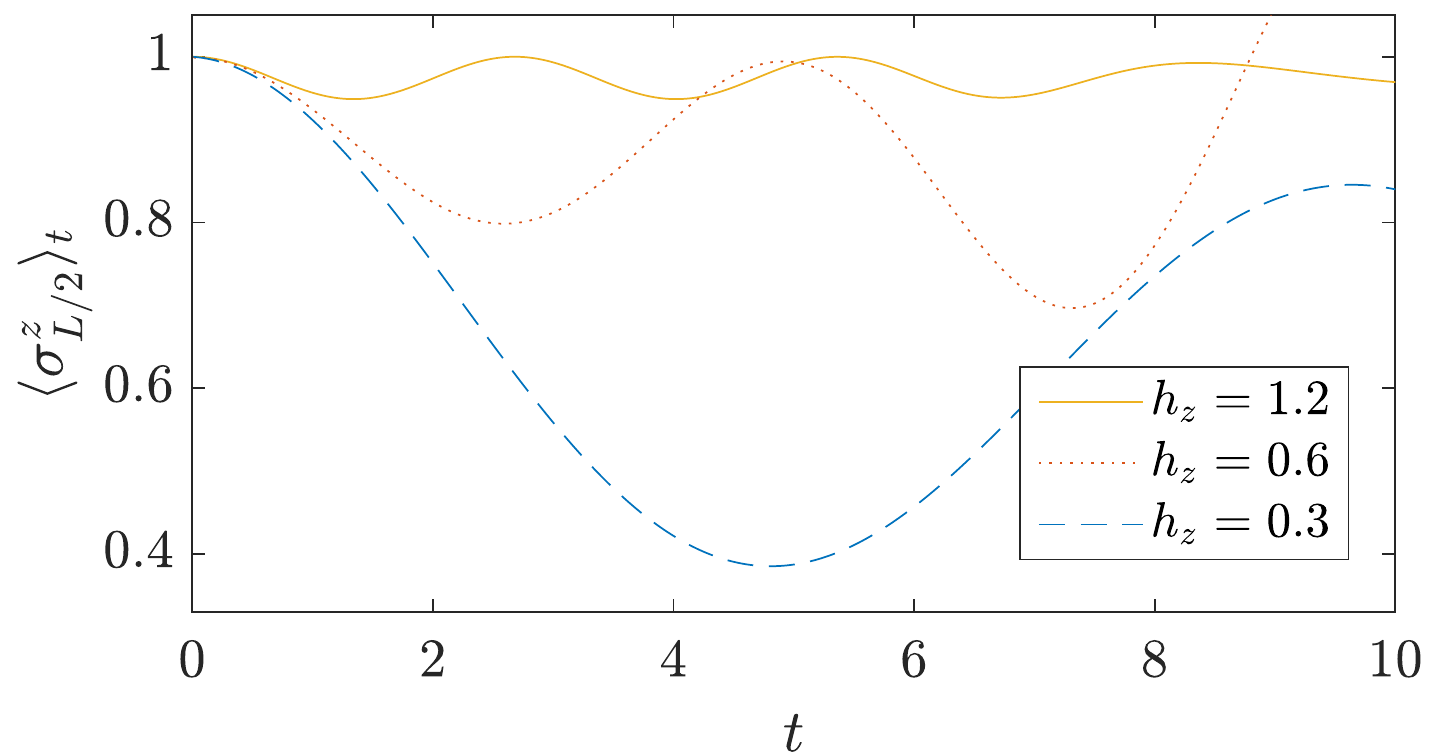}
    \caption{Time evolution of the magnetization $\langle \sigma_{L/2}^z\rangle_t$ calculated from an inverse Laplace transform from the data of Fig.\ \ref{fig:confinement_oscillations}.}
    \label{fig:timedomain}
\end{figure}

Using our algorithm, we now show how spin dynamics can be suppressed in an Ising chain in the above conditions. Similarly to the XXZ model studied in the previous subsection, we initialize the evolution of the spin chain from the domain wall state $|\text{ini}\rangle = |\uparrow \dots \uparrow\downarrow \dots \downarrow\rangle$. We quench the longitudinal field from $h_z = 0$ to a non-zero value and study the resulting dynamics. The initial state $|\text{ini}\rangle$ is characterized by a spin kink exactly in the middle of the chain. Hence, a non-zero longitudinal field will induce an energy imbalance between the two spin domains. Figure \ref{fig:energydens} shows the evolution in the Laplace domain of the energy density profile $\text{Tr}(H_i\tilde{\rho}_s)$, with
\begin{equation}
H_i = -J \sigma_i^z \sigma_{i+1}^z-\frac{h_x}{2}(\sigma_i^x + \sigma_{i+1}^x)-\frac{h_z}{2}(\sigma_i^z + \sigma_{i+1}^z),
\end{equation}
starting from the initial state Eq.\ \eqref{eq:inixxz} for an $L=40$ chain with $J=1$, $h_x = 0.2$, and $h_z = 1.2$. Neither energy exchange nor spin excitation propagation occurs between the two halves. This can be explained by considering the mid kink as a quasi-particle, whose motion is triggered by a non-zero longitudinal field $h_z$. The kinetic energy gain, which is of order $\sim h_z$, allows the kink to move within the potential, but due to energy conservation it has to periodically bounce back, leading to a confined dynamics (see \cite{confinement} for details). These oscillations are centered around the central bond connecting $L/2$ and $L/2+1$ and they can be measured by considering the evolution of $\langle\sigma^{z}_{L/2}\rangle$. This is illustrated in the upper panel of Fig.\ \ref{fig:confinement_oscillations} in the Laplace domain for a $L=20$ chain with $J=1$, $h_x=0.2$ and several confining potentials. Because the result is shown in the Laplace domain, the oscillations take the form of Lorezian curves.  
The mean frequency of the confining oscillations can be extracted as the inflection point of these curves. This is shown on the lower panel of Fig.\ \ref{fig:confinement_oscillations}, which illustrates the logarithmic derivative of the signal in the upper panel. The derivatives were calculated using a cubic spline fitting, and the oscillations frequency is represented by the location of the peak. Furthermore, the oscillation amplitudes can be estimated from the distance between the values $C_{s=0}^{\sigma_{L/2}^z}$  and $C_{s= \infty}^{\sigma_{L/2}^z}$. These amplitudes and frequencies scale as $\sim h_x/h_z$ and $\sim h_z$, respectively, which is in agreement with the quasi-particle interpretation of the kink dynamics and with the findings reported in \cite{confinement}.
This is further supported by Fig. \ref{fig:timedomain}, which displays the time evolution of the mean value of the magnetization $\langle \sigma_{L/2}^z\rangle_t$ obtained via an inverse Laplace transform of the data in the upper panel in Fig.\ \ref{fig:confinement_oscillations}, using the Zakian method. The inverting procedure relies on a rational polynomial fit of the original data in order to extrapolate the whole Laplace-transformed function. It starts by fitting the signal in Fig.\ \ref{fig:confinement_oscillations} by a function of the type $P_2/P'_2$ where $P_2$, $P'_2$ are polynomials of degree two. The optimal parameters of the first fit are then used as an initial guess for a fit of higher order polynomials, say $P_3/P'_3$. This iterative scheme continues until the parameters start diverging. As illustrated in Fig.\ \ref{fig:timedomain}, the frequency of the oscillations depends linearly on $h_z$ while the amplitude is inversely proportional. The numerical inverse Laplace method induces numerical amplitude damping, due to the use of finite of the $s$ range. The sharp increase in the magnetization for $h_z=0.6$ at about $t \approx 8$ comes from the dynamical sign problem as well as an interplay between the numerical inverse method and the extrapolation over the whole $s>0$ domain, suggesting that the results in the real time domain can be trusted up to $t \approx 5$. It is however worthy of noting that the Laplace signal does not contain any signs of divergences and can hence be trusted at least up to $s\approx 10$.

\section{Conclusions and open perspectives}
\label{concs}

In this work, we have introduced an alternate Monte Carlo method to study the reversible dynamics of many-body quantum systems. The validity and the accuracy of the method was benchmarked against excitation propagation in the Heisenberg XXZ model and dynamical confinement in the quantum Ising chain. In both cases, we were able to resolve long-time dynamical properties.

Similarly to FPQMC, this method is based on a piece-wise stochastic deterministic two-process unravelling to solve the von Neumann equation. The introduction of an importance sampling procedure allows us to limit the exploration of the Hilbert space to statistically important states. Furthermore, large times can be reached with the deadweight approximation by preventing statistically unimportant triplets from spawning. Due to the unravelling and the form of the evolution operator in the Laplace domain, these triplets can be kept in the simulation by only participating to the free evolution, which greatly reduces the statistical errors on the final result and delays the appearance of the divergences due to the dynamical sign problem. The trace of the density matrix is conserved at all times during the simulation. The maximum simulation time accessible is highly dependent on the observable, as it has been shown between the Loschmidt echo and the oscillations in the quantum Ising model. Unfortunately, we doubt that larger time scales can be reached. 

Regarding future directions, we believe that the introduction of a small dissipation parameter can reduce the severity of the sign problem, an idea that is already under current investigation. Thus, we are confident that our method can be extended to dissipative many-body quantum systems. 

\appendix

\section{Compression and decompression steps} \label{sec:comp}
Prior to the execution of the loop the ensemble is modified as to improve the statistics without influencing directly the averages. This modification is carried out through \textit{compressions} or \textit{decompressions}. In a compression, classes of triplets are formed by grouping together all the triplets associated to a fixed pair of states, for instance, $(i,j)$. These are then replaced by a single triplet whose weight is equal to the sum of the weights of all the members of the class.
Decompression is applied on a compressed ensemble. 
A single class of triplets $(i,j)$ is split into triplets with weight $w_u>0$ in absolute value,  $(\text{sgn}(w_n)w_u,i_n,j_n)$, and a single rest triplet $(w_r,i_n,j_n)$, with $w_r =\text{sgn}(w_n)(|w_n/w_u|-\lfloor |w_n/w_u|\rfloor)$ ($\lfloor \cdot \rfloor$ is the floor function). The rest triplet is then removed from the simulation with probability $1-|w_r|$; otherwise its weight is updated to $\text{sgn}(w_n)w_u$. That way, the total statistical weight is conserved on average.

\bibliography{biblio.bib}

\end{document}